# Anderson transition at complex energies in one-dimensional parity-time-symmetric disordered systems

Wei Wang[1,2]*, Xulong Wang[1]*, Guancong Ma[1†]

1. Department of Physics, Hong Kong Baptist University, Kowloon Tong, Hong Kong, China

2. School of Science, Harbin Institute of Technology, Shenzhen 518055, China

* These authors contributed equally to this work.

† Email: phgcma@hkbu.edu.hk

**Abstract**

The presence of disorder can severely impede wave transport, resulting in the famous Anderson localization. Previous theoretical studies found that Anderson transition can exist in one-dimensional (1D) non-Hermitian disordered rings with chiral hopping, defying the scaling theory of localization for Hermitian systems. In these systems, localized (extended) modes are associated with real (complex) energies. Here, we report that Anderson localized modes with complex energies can also exist in such systems. The emergence of the complex-energy localized modes (CELMs) directly ties to the properties of the corresponding pristine non-Hermitian system. Specifically, the density of states of the complex spectrum under the periodic boundary condition and the non-Bloch parity-time transition of the open-boundary chain both play critical roles in the emergence of the CELMs. The coexistence of extended modes, real-energy localized modes (RELMs), and CELMs should be a generic phenomenon for 1D non-Hermitian disordered systems under class AI. Our work shows that the interplay between Anderson mechanism and non-Hermitian physics enriches the properties of disordered media and opens new possibilities for controlling wave transport.

*Introduction.*— Anderson localization [1–3] is a long-standing research topic that has not only revolutionized the understanding of electron transport in solid-state materials [2,3] and the dynamics of quantum matters [4–6] but also opened new paradigms for electromagnetism, optics, photonics [7–11], and acoustics [12,13]. Yet Anderson localization has been traditionally studied under Hermitian frameworks. In the late 1990s, pioneering theoretical studies by Hatano and Nelson [14–16] investigated Anderson localization under non-Hermitian formalism, in which energy can be exchanged with external reservoirs [17–19]. The studies revealed that in the presence of nearest-neighbor chiral hopping, a one-dimensional (1D) disordered ring can simultaneously sustain both localized modes and extended modes, which are associated with real and complex energies, respectively [14,16,20–22]. This surprising discovery challenges the conclusion of the scaling theory of localization, which states that Anderson transition must be absent in 1D and all modes localized in the presence of disorder. This finding shows that non-Hermitian disordered systems have drastically different behaviors from their Hermitian counterparts. Rather surprisingly, experimental validation of Hatano and Nelson's prediction remains nonexistent to date.

Recently, non-Hermitian physics has witnessed tremendous developments in realms ranging from classical waves and optics [23,24] to condensed matter physics and quantum systems [19,25,26]. The distinction between complex conjugation and transposition for non-Hermitian Hamiltonian results in significant enrichment of topology, intrinsic symmetry, and universality classes [27]. Non-Hermitian spectral topology manifests as spectral loops (areas) for lattice systems, and it is closely related to non-Hermitian skin effects (NHSEs) [30–33]. Remarkably, NHSE can cause topological edge states to delocalize [34–36]. Such an effect is phenomenologically resemblance to the delocalization of Anderson localized modes predicted by Hatano and Nelson [14–16].

Here, we discover that Anderson transition can occur for complex-energy modes, which were thought to be exclusively extended modes in 1D non-Hermitian disordered rings. As such, RELMs, CELMs, and extended modes can exist in a single lattice at different energies. The emergence of CELMs can be straightforwardly seen from the peaks of the density of states (DOS) of the pristine model under the periodic boundary condition (PBC). We also reveal that the CELMs are fundamentally related to the non-Bloch parity-time ($\mathcal{PT}$) transition [37,38] of the pristine model under the open boundary condition (OBC). The coexistence of RELMs, CELMs, and extended modes is generic for 1D systems under non-Hermitian class AI. By using active mechanical lattices, we have not only experimentally validated Hatano and Nelson's predictions but also successfully observed the CELMs.

***Theoretical model.***— We start with a 1D non-Hermitian ring [Fig. 1(a)] with a Hamiltonian

$$\mathcal{H}_p = \sum_{j=1}^{L}\bigl(tc_{j+1}^{\dagger}c_j + \text{h.c.}\bigr) + \sum_{j=1}^{L}\bigl(\delta_1 c_{j+1}^{\dagger}c_j + \delta_2 c_{j+2}^{\dagger}c_j\bigr), \quad (1)$$

where $c_j^{\dagger}$ ($c_j$) is the creation (annihilation) operator with $j$ indexing the sites and $L$ being the total number of sites. Model (1) is denoted as a pristine model. We set $L = 400$ in all theoretical results in this work. The wavefunctions $\psi(x_j)$ satisfy the Born von-Karman boundary condition (BKBC), i.e., $\psi(x_j) = \psi(x_j + L)$. All parameters in Eq. (1) are real. $t$ is the nearest-neighbor hopping. $\delta_{1,2}$ are the nearest and second-neighbor chiral (or non-reciprocal) hopping, respectively. More general configurations, *e.g.*, third-neighbor non-reciprocity, are explored in ref. [39]. Equation (1) obeys $\mathcal{H}_p = \mathcal{H}_p^*$ and $\mathcal{H}_p \neq \mathcal{H}_p^T$, so it belongs to class AI in non-Hermitian universality classes [27,40]. (It also respects generalized $\mathcal{PT}$-symmetry, *i.e.*, $\mathcal{CH}_p\mathcal{C} = \mathcal{H}_p$, where $\mathcal{C}$ is the complex conjugate operator.) Consequently, the spectrum of Eq. (1) is either real or in complex conjugate pairs. Because Eq. (1) is periodic, its spectrum, provided $L$ is sufficiently large, matches that of Bloch Hamiltonian under the PBC

$$H_p(k) = 2t\cos k + \delta_1 e^{-ik} + \delta_2 e^{-2ik}, \quad (2)$$

where $k$ is the Bloch wavevector. In Figs. 1(b, e), we plot the spectra of Eq. (2) in two different cases, each with only one non-vanishing chiral hopping. The Bloch spectra form different loops on the complex-energy plane. Therein, the blue color scale represents the DOS, given by $D(E) = \frac{1}{2\pi}\sum |dH_p(k)/dk|^{-1}$, where the summation runs over all $k$ that satisfies $E = H_p(k)$, which is necessary because different $k$ may be mapped to the same energy, *i.e.*, the spectra may have crossings on the complex-energy plane. The DOS exhibits peaks when the spectral loops cross the real axes. Upon the introduction of disorder, localization shall first emerge from the positions where the DOS is maximum.

***Emergence of CELMs in disordered rings.***— Next, we introduce disorder as real random potential into the system, and the disordered Hamiltonian reads

$$\mathcal{H}_d = \mathcal{H}_p + \sum_{j=1}^{L} V_j c_j^{\dagger}c_j, \quad (3)$$

where $V_j = W \times \text{rand}(-1,1)$ is the real-valued disorder potential. Note that Eq. (3) still belongs to class AI and the generalized $\mathcal{PT}$-symmetry remains intact. We examine the spectra and eigenmodes of Eq. (3) across three different scenarios. Case-I has $\delta_1 \neq 0$ and $\delta_2 = 0$ [Fig. 1(c)]. A pair of "wings" lying on the real axis develop from the spectral loop. (This resembles the case previously studied by Hatano and Nelson [14–16].) We show that the level spacing in the left wing conforms to the Poisson distribution [39], a signature of Anderson localization [41], suggesting that modes on the wings are localized in character. This is confirmed by examining the wavefunctions [Fig. 1(d)]. An alternative way to characterize the localization is by the biorthogonal inverse participation ratio (BIPR) [42], defined as $R_m = \frac{\sum_{j=1}^{L} \rho_{m,j}^2}{\left(\sum_{j=1}^{L} \rho_{m,j}\right)^2}$, where $m$ labels the eigenmode, $\rho_{m,j} = |\psi_{m,j}^L \psi_{m,j}^R|$ denotes the biorthogonal density, and $\psi_{m,j}^L$, $\psi_{m,j}^R$ are the entries of the left and right eigenvectors at site-$j$, respectively. It is evident that real-energy modes (exhibiting large BIPRs) are localized, while complex-energy modes are extended, aligning with previous investigations on 1D non-Hermitian disordered rings [14,16,20–22]. Similar phenomena are also seen in case-II with $\delta_1 = 0$, $\delta_2 \neq 0$ [Figs. 1(e, f)]. Here, the spectrum of $H_\text{p}(k)$ has two loops, but the energy of localized modes remains real.

However, the association of real energy and localization can be broken and localized modes with complex energies emerge. To show this, we examine case-III, with $\delta_1 = 0$ and $\delta_2 = -0.3$ (Fig. 2). Comparing Fig. 2(a) to Fig. 1(e), the difference in the pristine models is simply an increase in $|\delta_2|$, which slightly deforms the double-loop Bloch spectrum. Yet the change in $\delta_2$ clearly also affects the DOS, which no longer peaks at the real axis. Localized modes are, therefore, unlikely to first appear around the real axis. As shown in Fig. 2(b), they indeed first appear in the proximity of the DOS maxima as scattered points on the complex plane. Their localized nature is further confirmed by examining their real-space wavefunctions and their BIPRs. Note that the CELMs are not special cases, they appear in massive numbers with sufficient disorder [Figs. 2(c, d)]. The level spacings of both the real and imaginary parts of the CELMs follow Poisson distribution [39].

We further analyze the DOS as functions of $k$ for different $\delta_2$. In Fig. 3(a), the DOS is dominated by three peaks and two small spikes. A peak is seen at $k = 0$, where the left spectral loop of the PBC spectrum intersects with the real axis. The small spikes at $k = \pm\pi/2$ correspond to the crossing point of the spectra where the number of states is doubled. Two additional peaks are seen near $k = \pm\pi$ and they become van Hove singularities at $k = \pm 0.75\pi$ as $|\delta_2|$ increases to 0.4. As localization typically initiates at DOS maxima, the increase in $|\delta_2|$ then triggers the emergence of the CELMs. This is in stark contrast with the Hatano-Nelson model [14–16], wherein disorder and $|\delta_1|$ are two competing factors, and the increase of the latter only causes more eigenmodes to delocalize.

From Figs. 1(f) and 2(c), it can be inferred that there exists a critical strength for the emergence of CELMs. To investigate this, we analyze the number of CELMs (denoted as $N_{\text{CL}}$) as a function of $|\delta_2|$ [upper panel in Fig. 3(b)]. As shown, CELMs begin to appear when $|\delta_2| > |\delta_2^c| = |t/3\sqrt{3}|$. In the next section, we show that this critical value coincides with the non-Bloch $\mathcal{PT}$-symmetry phase transition point in the pristine open chain [lower panel in Fig. 3(b)].

Further, we study the number of localized modes (denoted $N_{\text{L}}$, including both the real-energy and complex-energy ones) as a function of $\delta_2$. The presence of small $|\delta_2|$ first causes delocalization ($N_{\text{L}}$ decreases), but then $N_{\text{L}}$ increases and then peaks around $\delta_2 = -0.4$. The color of the curve in Fig. 3(c), representing the percentage of the CELMs, distinctly indicates that the CELMs become dominant when $\delta_2 < -0.2$ for a given $W$.

***Role of non-Bloch $\mathcal{PT}$-broken phase.***— Non-Bloch $\mathcal{PT}$ transition also plays a critical role in the emergence of CELMs [Fig. 4(a)]. To illustrate this, consider the fact that localized modes can only sense what happens far from the localization center through their decaying tails. It follows that we can cut the ring to an open chain without affecting the wavefunctions and energies of most localized modes [43]. Take case-II [Fig. 1(f)] as an example. Upon converting to OBC by cutting the ring at site $j = 1$, all modes have real energies. Specifically, the extended modes under the BKBC become skin modes with real energies [Fig. 4(b)]. This can be better seen by defining the localization center as $\chi_m = \sum_{j=1}^{L} j |\psi_{m,j}^R|^2$. Skin modes are likely to have $\chi_m \to 1$ or $\chi_m \to L$. For our lattice with $L = 400$, we set the skin-mode thresholds to be $\chi_m \leq 20$ and $\chi_m \geq 380$. In Fig. 4(b), it is seen that most of such modes fall within the loops of the BKBC spectrum, confirming that they are indeed mostly skin modes. In comparison, the modes with $\chi_m$ in between the thresholds have energies largely overlapping with the BKBC spectrum. They are mostly Anderson localized modes [44] and their energies are not affected by the change in the boundary condition. The results for case-I are presented in [39], where similar phenomena are observed.

Now consider case-III with $\delta_2 = -0.3$ [Fig. 2(c)]. In Fig. 4(d), under the OBC, the $\chi_m$ thresholds sieve out skin modes developed from the two loops of the BKBC spectrum. However, only the energies of the skin modes from the left loop collapse to the real axis. The ones from the right loop remain scattered on the complex plane, which again differs from case-II. The reason for these differences is revealed by a simple look at the pristine open chains: case-II features a purely real spectrum [green dots in Fig. 4(c)], while case-III is complex-valued in the right part of the spectrum [Fig. 4(e)]. In other words, from case-II to case-III, the pristine OBC system has undergone a non-Bloch $\mathcal{PT}$ transition [37,45] caused by the increase in non-Hermiticity ($|\delta_2|$). With the formula derived in ref. [45], we can get the critical values, *i.e.*, $|\delta_2^c| = |t/3\sqrt{3}|$. Accordingly, the pristine OBC system is in the non-Bloch $\mathcal{PT}$-broken phase when $|\delta_2| > |t/3\sqrt{3}|$ for $\delta_1 = 0$ [lower panel in Fig. 3(b)]. Consequently, the emergence of CELMs can be understood as the effect of real random potentials on non-Bloch $\mathcal{PT}$-broken modes with complex energies. Such connection becomes more apparent when examining how it evolves with $W$ [39].

*Experimental realizations.*— We use an active mechanical lattice comprising 60 rotational oscillators [34,35] to realize the three different cases and to observe the CELMs. In the experiments, we measure the responses of the system under the single-site excitation to characterize the localized and extended modes. A harmonic torque at a chosen frequency is applied to an oscillator by driving the motor and the angular displacements at all oscillators are recorded. To fully characterize the system, the procedure is then repeated with each oscillator excited individually. In the first experiment, the lattice is set to $t = -0.57$, $\delta_1 = -0.4$, and $W = 0.95$ [46]. Figure 5(a) shows the computed complex spectra of 25 independent configurations of random potentials. The eigenmodes around 8 Hz and 10 Hz are on the left wing and the loop in the spectrum. The lattice is then excited at these two frequencies respectively [marked by arrows in Fig. 5(a)]. The configuration-averaged responses (25 realizations) are shown in Figs. 5(b, c), wherein the horizontal (vertical) axis represents the excitation (probe) position. It is seen that, at 8 Hz, the results give an anti-diagonal distribution [Fig. 5(b)], which means only the excitation positions have large oscillation amplitudes. In other words, the modes are localized. In contrast, the responses at 10 Hz show that the wave can propagate in the lattice–clear evidence of the existence of extended modes [Fig. 5(c)]. The unidirectional propagation conforms with the system's non-reciprocal nature and the eventual decay is due to the dissipation. These results are the experimental demonstration of Hatano and Nelson's prediction [14–16].

Next, we experiment with the other two cases ($\delta_1 = 0$, $\delta_2 \neq 0$). Figure 5(d) shows the spectra of 25 configurations of disorder with $t = -0.57$, $\delta_2 = -0.34$, and $W = 0.44$. [The OBC spectrum is in the non-Bloch $\mathcal{PT}$-broken phase (Fig. S23).] From the BIPRs, we can identify the presence of the CELMs around the real frequency of 10.8 Hz. We then measure the response at 9.2 Hz and 10.8 Hz [Figs. 5(e, f)]. At 9.2 Hz, the responses clearly show propagative characteristics. But at 10.8 Hz, the responses are strongly confined at the excitation positions, confirming the existence of the CELMs. As a controlled experiment, we further set $t = -2.06, \delta_2 = -0.33$ [in the non-Bloch $\mathcal{PT}$-exact phase (Fig. S24)], and $W = 0.46$ [Figs. 5(g-i)]. The results confirm the absence of the CELM. We have experimentally measured the case of the third-neighbor chiral hopping, in which the CELMs also exist [39].

*Discussion.*— The CELMs reported here are nontrivial due to their fundamental links to spectral topology and non-Bloch $\mathcal{PT}$-transition of the pristine models. The coexistence of RELMs, CELMs, and extended modes indicates that Anderson transition can occur on the complex energy plane in 1D non-Hermitian disordered systems. Such transition generically exists for 1D non-Hermitian disordered rings under class AI. Several additional numerical examples are presented in the ref. [39] to support our conclusion. Note that CELMs alone can "trivially" emerge in other non-Hermitian universality classes, but not in such a way that they coexist with both RELMs and extended modes. To be specific, we can consider two different cases [39]. In the first case, we simply introduce complex onsite disorders to Eq. (1). This changes the system to class A, since the Hamiltonian now satisfies $\mathcal{H}_d \neq \mathcal{H}_d^*$ and $\mathcal{H}_d \neq \mathcal{H}_d^T$. CELMs trivially emerge because of the imaginary onsite entries, but RELMs are generically absent. In the second case, consider a reciprocal system with onsite gain and loss, then $\mathcal{H}_d \neq \mathcal{H}_d^*$ and $\mathcal{H}_d = \mathcal{H}_d^T$, so the system falls into class AI[†]. Now, all modes are localized and Andeson transition at complex energies is absent.

Summarizing our findings, we can arrive at a simple recipe for designing localized modes in 1D disordered rings in class AI. To predict the existence of the CELMs, one merely needs to check whether the pristine OBC lattice is in the non-Bloch $\mathcal{PT}$-broken phase and then provide a sufficiently strong onsite disorder. Then, the DOS, which is directly obtainable from the pristine PBC spectrum, is informative about the exact spectral position of the onset of the Anderson localization upon introducing real random potential (We show in ref. [39] that by tuning the chiral hoping, the sequence in which the RELMs and CELMs appear as $W$ increases can be further controlled.). On the contrary, if the goal is to guarantee only RELMs exist, a reliable way is to ensure that the corresponding OBC spectrum is in the non-Bloch $\mathcal{PT}$-exact phase.

CELMs open new possibilities for the exploration of Anderson localization. When the CELMs appear in massive amounts, they can dominate the properties of the system and entail novel transport phenomena, which are interesting topics for future studies. The CELMs are also expected to find useful applications. For example, half of the CELMs are temporally divergent gain modes, making them interesting for applications requiring amplifications, such as sensors and lasers [47–49], whereas the loss modes may be useful for total absorption [50].

*Acknowledgment.*— G.M. thanks Zhao-Qing Zhang and Ping Sheng for fruitful discussions. This work was supported by the National Key R&D Program (2022YFA1404403), the Hong Kong Research Grants Council (RFS2223-2S01, 12301822, 12302420), and the National Natural Science Foundation of China (12472088).

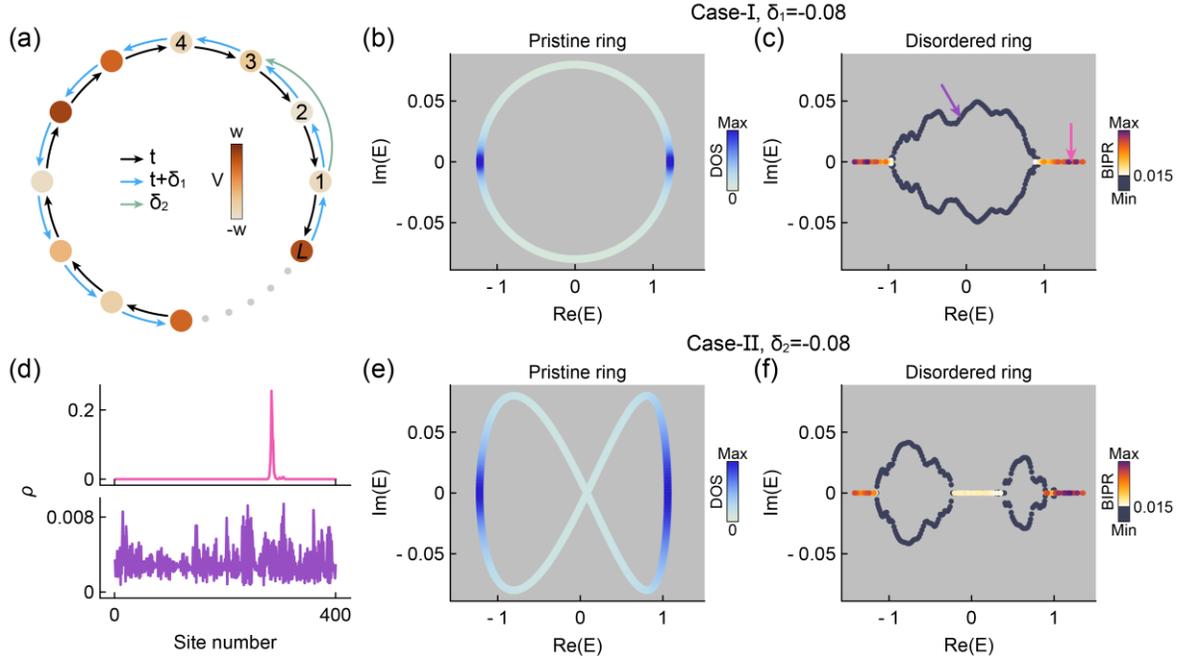

FIG. 1 (a) The schematic model of a non-Hermitian disordered ring. (b, c, e, f) The complex spectra of (b, e) the pristine ring and (c, f) the disordered ring. The blue and yellow color scales represent the DOS and the BIPR, respectively. Modes with a BIPR smaller than 0.015 are extended in character and are plotted in black. (d) The biorthogonal density $\rho$ of the two eigenmodes marked by the arrows in (c). In all calculations, $t = -0.57$ and $W = 0.4$.

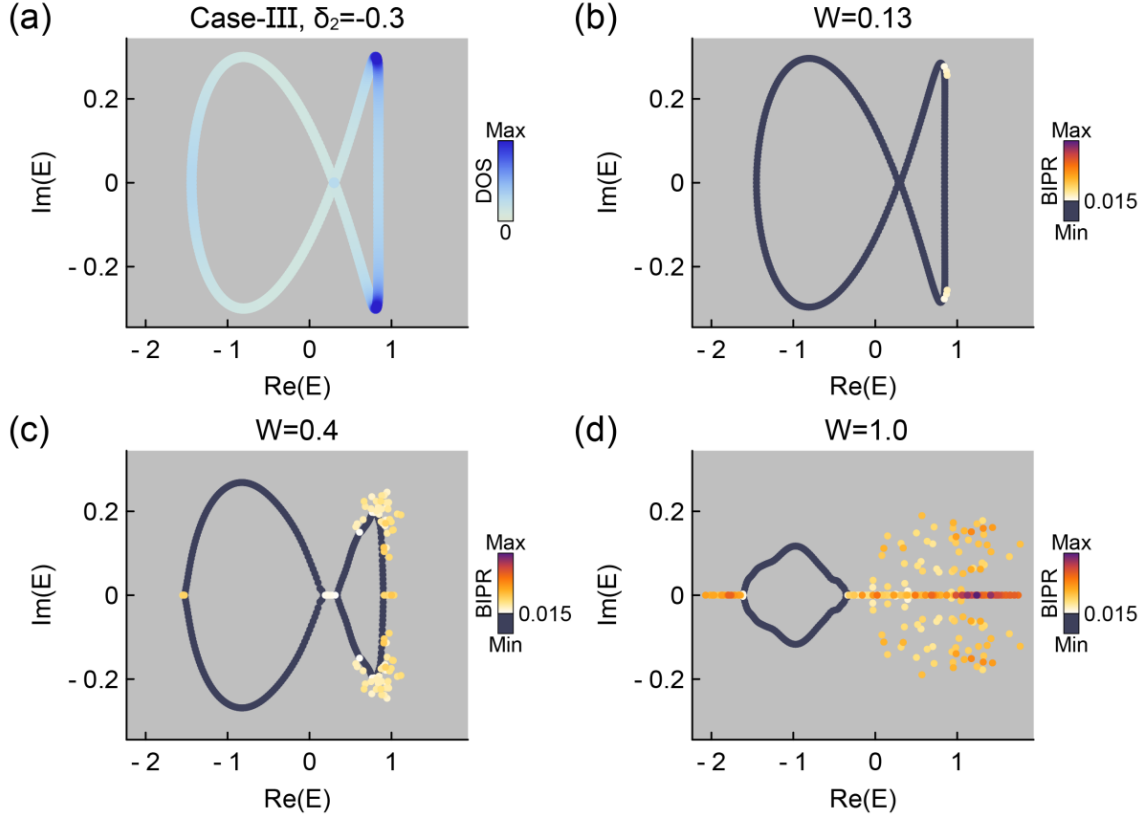

FIG. 2 (a) The complex spectrum of the pristine ring at $\delta_2 = -0.3$ with the blue color scale representing the DOS. (b-d) The evolution of the complex spectrum of the disordered ring when linearly scaling real random onsite potential. The yellow color scale represents the BIPR, with modes associated with a BIPR smaller than 0.015 plotted in black. In the calculations, we set $t = -0.57$, (b) $W = 0.13$, (c) $W = 0.4$, and (d) $W = 1.0$.

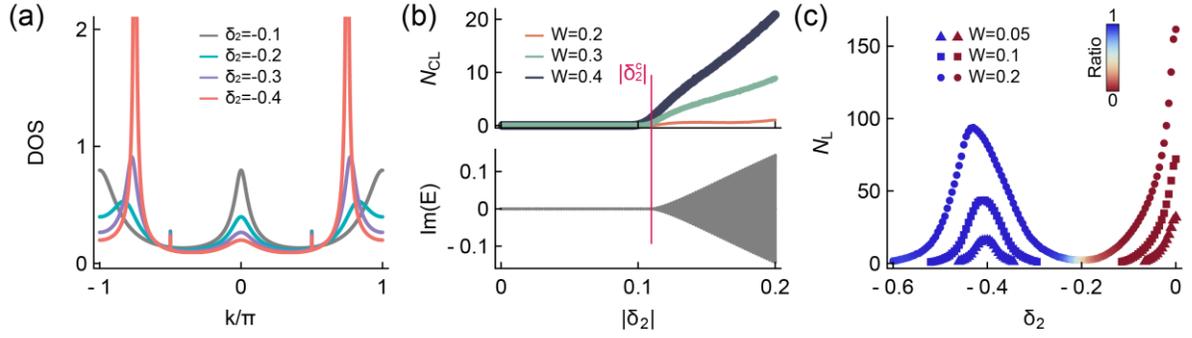

FIG. 3 (a) The DOS as functions of $k$ at different $\delta_2$. (b) The number of the CELMs $N_{\mathrm{CL}}$ as functions of $|\delta_2|$ at different $W$ (upper panel). The imaginary parts of the energy spectra of the pristine open chain as a function of $|\delta_2|$ (lower panel). (c) The number of localized modes $N_{\mathrm{L}}$ as functions of $\delta_2$ at different $W$. The color scale represents the ratio between the number of CELMs and $N_{\mathrm{L}}$. In (b, c), the results are averaged over 5000 realizations of onsite disorder.

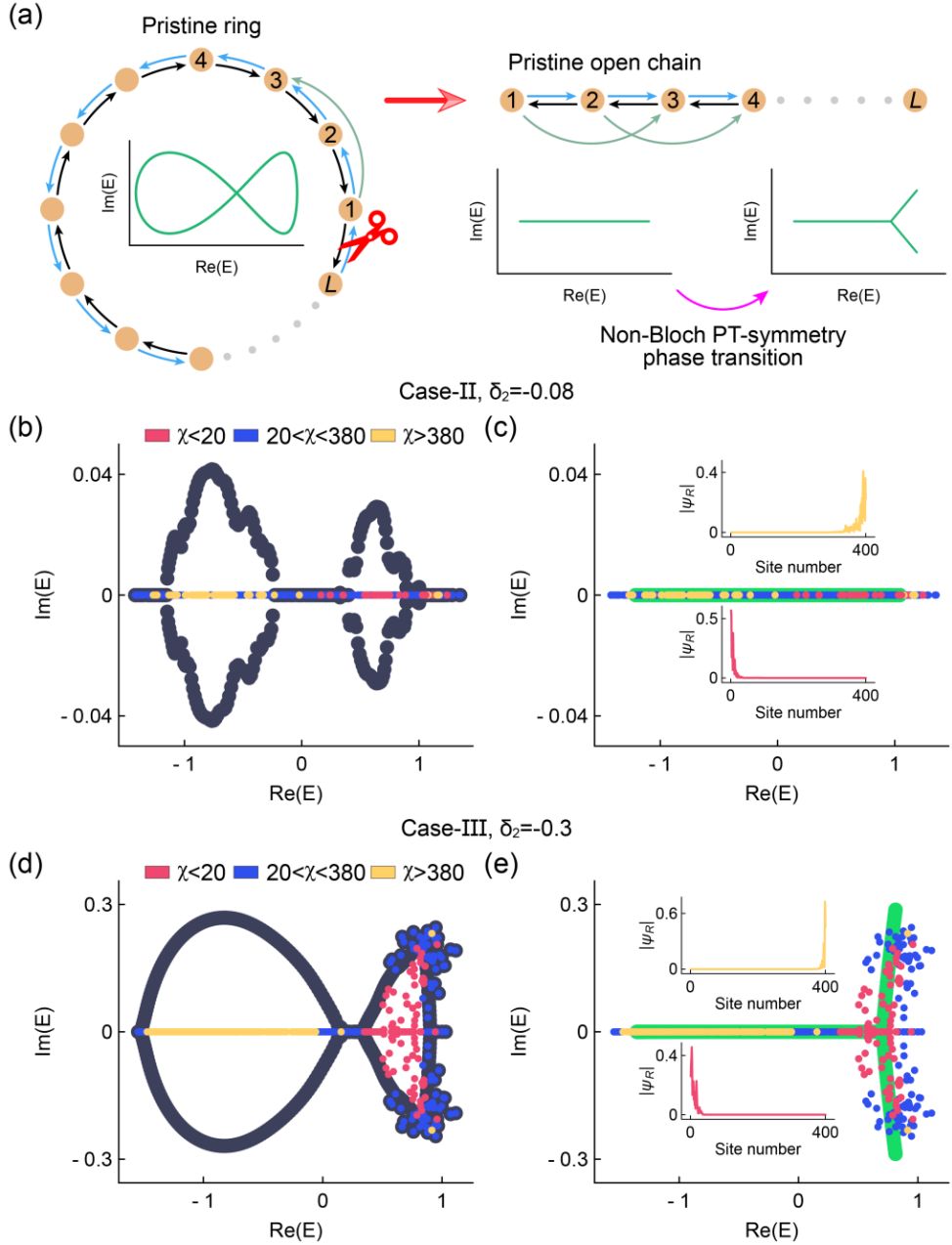

FIG. 4 (a) An open chain is formed by cutting the ring. Non-Bloch $\mathcal{PT}$ transition can occur in the pristine open chain [lower panel in Fig. 3(b)]. (b, d) The black dots plot the spectra of the disordered ring of (b) case-II and (d) case-III, respectively. (c, e) The green dots are the spectra of the corresponding pristine open chain. The pink, blue, and yellow dots in (b-d) denote the spectra of the corresponding disordered open chain, with colors distinguishing the localization center $\chi_m$. The insets in (c, e) show two skin-mode wavefunctions localized at the left (pink) and right (yellow) end of the disordered open chain. $W = 0.4$ in all results.

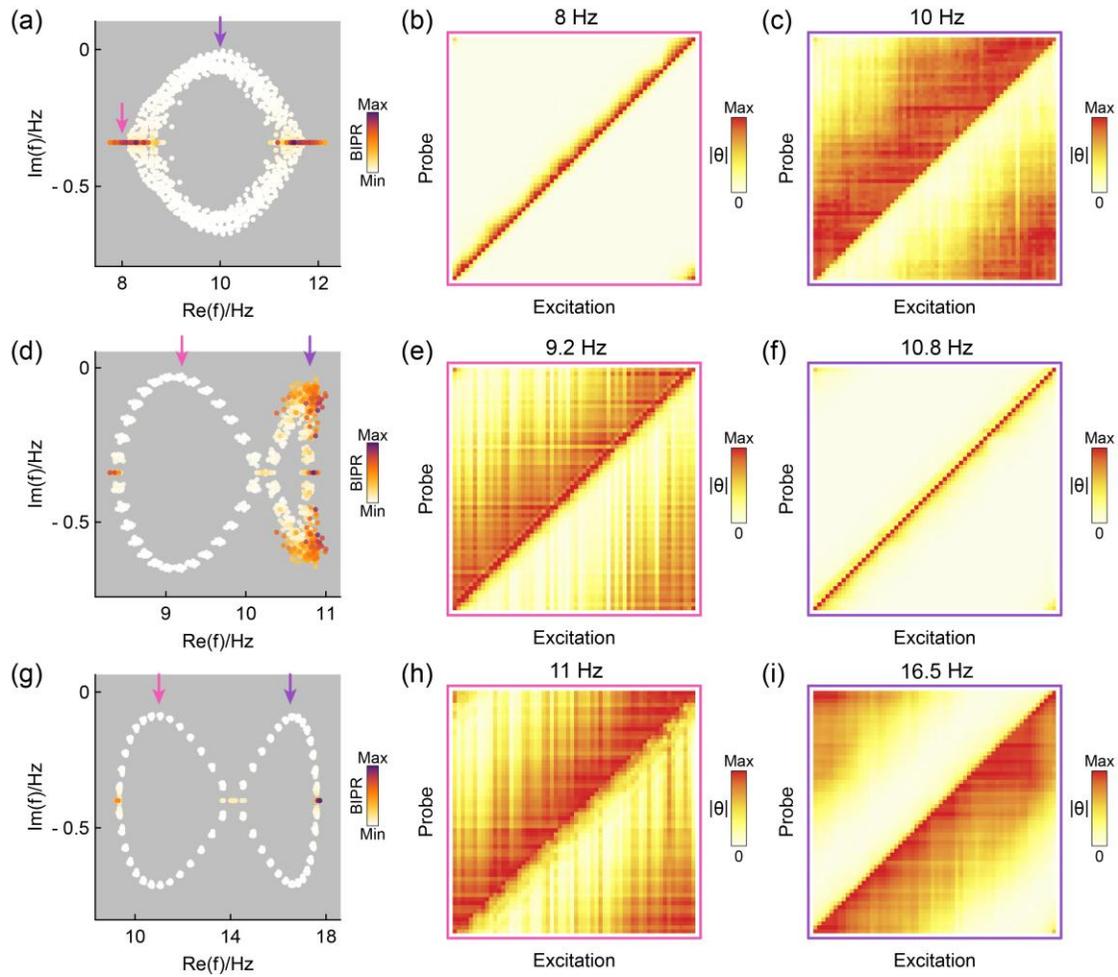

FIG. 5 (a, c, e) The complex spectra of the disordered ring with 25 realizations of real onsite disorder. The yellow color represents the BIPR. (b, c, e, f, h, i) The response maps at selected frequencies. The experimental parameters used in (a, d, g) are listed in [39].